\documentstyle[prl,aps,floats,epsf,psfig,colordvi]{revtex}
\draft
\newcommand{\nub}{\overline{\nu}}
\newcommand{\sbar}{\overline{s}}

\def\nim#1#2#3  {{\em Nucl. Instr. Meth.} {\bf#1}, #2 (#3). }
\def\np#1#2#3   {{ Nucl. Phys.} {\bf#1}, #2 (#3). }
\def\pcps#1#2#3 {{ Proc. Cam. Phil. Soc.} {\bf#1}, #2 (#3). }
\def\pl#1#2#3   {{ Phys. Lett.} {\bf#1}, #2 (#3). }
\def\plc#1#2#3   {{ Phys. Lett.} {\bf#1}, #2 (#3); }
\def\prep#1#2#3 {{ Phys. Rep.} {\bf#1}, #2 (#3). }
\def\prev#1#2#3 {{ Phys. Rev.} {\bf#1}, #2 (#3). }
\def\prl#1#2#3  {{ Phys. Rev. Lett.} {\bf#1}, #2 (#3). }
\def\prs#1#2#3  {{ Proc. Roy. Soc.} {\bf#1}, #2 (#3). }
\def\ptp#1#2#3  {{ Prog. Th. Phys.} {\bf#1}, #2 (#3). }
\def\rmp#1#2#3  {{ Rev. Mod. Phys.} {\bf#1}, #2 (#3). }
\def\rpp#1#2#3  {{ Rep. Prog. Phys.} {\bf#1}, #2 (#3). }
\def\zp#1#2#3   {{ Z. Phys.} {\bf#1}, #2 (#3). }
\def\epj#1#2#3   {{ Eur. Phys. Jour.} {\bf#1}, #2 (#3). }

\begin{document}

\wideabs{
\title{ Extraction of  $R=\frac{\sigma_L}{\sigma _T} $ from CCFR
     $\nu_\mu$-Fe and $\nub_\mu$-Fe differential cross sections}

\author{U.~K.~Yang,$^{7}$ T.~Adams,$^{4}$ A.~Alton,$^{4}$
C.~G.~Arroyo,$^{2}$ S.~Avvakumov,$^{7}$ L.~de~Barbaro,$^{5}$
P.~de~Barbaro,$^{7}$ A.~O.~Bazarko,$^{2}$ R.~H.~Bernstein,$^{3}$
A.~Bodek,$^{7}$ T.~Bolton,$^{4}$ J.~Brau,$^{6}$ D.~Buchholz,$^{5}$
H.~Budd,$^{7}$ L.~Bugel,$^{3}$ J.~Conrad,$^{2}$ R.~B.~Drucker,$^{6}$
B.~T.~Fleming,$^{2}$
J.~A.~Formaggio,$^{2}$ R.~Frey,$^{6}$ J.~Goldman,$^{4}$
M.~Goncharov,$^{4}$ D.~A.~Harris,$ ^{7} $ R.~A.~Johnson,$^{1}$
J.~H.~Kim,$^{2}$ B.~J.~King,$^{2}$ T.~Kinnel,$^{8}$
S.~Koutsoliotas,$^{2}$ M.~J.~Lamm,$^{3}$ W.~Marsh,$^{3}$
D.~Mason,$^{6}$ K.~S.~McFarland, $^{7}$ C.~McNulty,$^{2}$
S.~R.~Mishra,$^{2}$ D.~Naples,$^{4}$  P.~Nienaber,$^{3}$
A.~Romosan,$^{2}$ W.~K.~Sakumoto,$^{7}$ H.~Schellman,$^{5}$
F.~J.~Sciulli,$^{2}$ W.~G.~Seligman,$^{2}$ M.~H.~Shaevitz,$^{2}$
W.~H.~Smith,$^{8}$ P.~Spentzouris, $^{2}$ E.~G.~Stern,$^{2}$
N.~Suwonjandee,$^{1}$ A.~Vaitaitis,$^{2}$ M.~Vakili,$^{1}$
J.~Yu,$^{3}$ G.~P.~Zeller,$^{5}$ and E.~D.~Zimmerman$^{2}$}

\address{( The CCFR/NuTeV Collaboration ) \\
$^{1}$ University of Cincinnati, Cincinnati, OH 45221 \\
$^{2}$ Columbia University, New York, NY 10027 \\
$^{3}$ Fermi National Accelerator Laboratory, Batavia, IL 60510 \\
$^{4}$ Kansas State University, Manhattan, KS 66506 \\
$^{5}$ Northwestern University, Evanston, IL 60208 \\
$^{6}$ University of Oregon, Eugene, OR 97403 \\
$^{7}$ University of Rochester, Rochester, NY 14627 \\
$^{8}$ University of Wisconsin, Madison, WI 53706\\ }

\date{\today}
\maketitle
\begin{abstract}

We report on the extraction of $R=\frac{\sigma_L}{\sigma _T} $
from CCFR  $\nu_\mu$-Fe and $\nub_\mu$-Fe
differential cross sections.
The CCFR differential cross sections do not show the deviations
from the QCD expectations that are seen in the CDHSW data at
very low and very high $x$.
$R$ as measured in $\nu_\mu$ scattering is in agreement
with $R$ as measured in muon  and electron scattering.
All data on $R$ for $Q^2 > 1$ GeV$^2$
are in agreement with a NNLO QCD calculation which uses NNLO PDFs and includes
target mass effects.
We report on the first measurements of  $R$
in the low $x$ and  $Q^2 < 1$ GeV$^2$ region (where an anomalous large
rise in $R$ for nuclear targets has been observed by the
HERMES collaboration).
\end{abstract}
\pacs{PACS numbers:12.38.Qk, 13.15.+g, 24.85.+p, 25.30.Pt
\twocolumn
}}


      The ratio of longitudinal and transverse structure function,
     $R$ (=$F_L/2xF_1$) in deep
     inelastic lepton-nucleon scattering experiments is a sensitive
test of the quark parton model of the nucleon. In leading order QCD,
$R$ for the scattering from spin 1/2
constituents (e.g. quarks) is zero, while $R$ for the scattering
from spin 0 or spin 1 constituents is very large. The small value
of $R$ originally measured in electron scattering experiments~\cite{BODEK}
provided the initial evidence for the spin 1/2 nature
of the nucleon constituents. However, a non-zero value of $R$
can also originate from processes in which the struck quark has
a finite transverse momentum. These include Quantum Chromodynamics
(QCD) processes involving
emissions of gluons, processes involving the production of
heavy quarks, target mass~\cite{GP} corrections and higher twist
effects~\cite{dupaper,YANGR}. Recently, there has been a renewed
interest in $R$ at small values of $x$ and $Q^2$, because of the large
anomalous nuclear effect that has been reported by the
HERMES experiment~\cite{HERMES}. A large value of $R$ in nuclear
targets could be interpreted as evidence for non spin 1/2
constituents, such as $\rho$ mesons in nuclei~\cite{MIT}.
In this letter, we report on an extraction of $R$
in neutrino scattering ($R^\nu$), extending to low $x$ and $Q^2$.
We also compare the
CCFR differential cross sections with
previous CDHSW~\cite{CDHS} $\nu_\mu$-Fe and $\nub_\mu$-Fe data.

Previous measurements of $R$ in muon and electron scattering ($R^{\mu/e}$)
were fit using $R_{world}^{\mu/e}$~\cite{RWORLD} (a QCD inspired
empirical form).
The $R_{world}^{\mu/e}$ fit is also in good agreement
with recent NMC muon data~\cite{NMCR} for $R$ at low $x$, and
with theoretical predictions~\cite{YANGR}
     $R_{NNLO+TM}^{\mu/e}$
     (a Next to Next to Leading (NNLO) QCD calculation using NLO
     Parton Distribution Functions (PDFs), and including target mass effects).
Very recently the NNLO-QCD calculations for $F_{L}$ have been
updated~\cite{MRSNNLO}
to include estimates of the contribution from NNLO PDFs.
In addition, the NLO-QCD calculations have been updated to include
    $ln(1/x)$ resummation terms~\cite{resum} which are important at
    small $x$. A full QCD calculation which
includes both the NNLO  and the $ln(1/x)$ resummation
terms is not yet available.  We evaluate
$R_{NNLOpdfs+TM}^{\mu/e}$ and  $R_{NLOresum+TM}^{\mu/e}$
by adding target mass effects
to these calculations of $F_{L}$.

For $x>0.1$ it is expected that $R^\nu$  should
be the same as $R^{\mu/e}$.
However, for $x<0.1$ and low $Q^2$  (in leading order),
$R^\nu$  is expected to be larger
than $R^{\mu/e}$ because of the production of
     massive charm quarks in the final state.
We calculate~\cite{WWW}
a correction to $R_{world}^{\mu/e}$ for this
difference  using a leading order slow rescaling model with a charm
mass, $m_c (=1.3$ GeV)
and obtain an  effective
$R_{world}$ for $\nu_\mu$ scattering ($R_{eff}^{\nu}$).  Our
measurements of $R^\nu$  are compared
to $R^{\mu/e}$ data and also to predictions from $R_{eff}^{\nu}$,
$R_{world}^{\mu/e}$, $R_{NNLOpdfs+TM}^{\mu/e}$,
and $R_{NLOresum+TM}^{\mu/e}$ .

Values of $R$ are extracted from
the sum of $\nu_\mu$ and $\nub_\mu$
     differential cross sections
for charged current interactions on isoscalar target
using following relation:
\begin{tabbing}
$F(\epsilon)$ \= $\equiv \left[\frac{d^2\sigma^{\nu }}{dxdy}+
\frac{d^2\sigma^{\overline \nu}}{dxdy} \right]
     \frac {(1-\epsilon)\pi}{y^2G_F^2M E_\nu }$ \\
      \> $ = 2xF_1 [ 1+\epsilon R ] + \frac {y(1-y/2)}{1+(1-y)^2} \Delta
xF_3 $, \hspace{0.7in} (1)
\end{tabbing}
where $G_{F}$ is the weak Fermi coupling constant, $M$ is the nucleon
mass, $E_{\nu}$ is the incident energy, the scaling
variable $y=E_h/E_\nu$ is the fractional energy transferred to
the hadronic vertex, $E_h$ is the final state hadronic
energy, and $\epsilon\simeq2(1-y)/(1+(1-y)^2)$ is the polarization of virtual
$W$ boson. The structure
function $2xF_1$ is expressed in terms of $F_2$
by $2xF_1(x,Q^2)=F_2(x,Q^2)\times
\frac{1+4M^2x^2/Q^2}{1+R(x,Q^2)}$, where $Q^2$ is the
square of the four-momentum transfer to the nucleon,
      $x=Q^2/2ME_h$ (the Bjorken scaling variable) is
the fractional momentum carried by the struck quark.
Here  $\Delta xF_3=xF_3^{\nu}-xF_3^{\nub}$, which in leading order
     $\simeq4x(s-c)$ (difference between the strange and charm
quark distributions).


\begin{figure}[bh]
\centerline{\psfig{figure=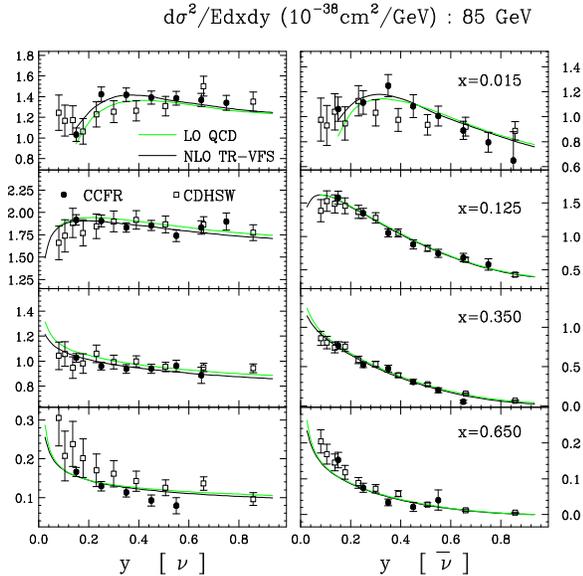,width=3.0in,height=3.0in}}
\caption{Some of the CCFR and CDHSW
differential cross section data at $E_\nu=85$
    (both statistical and systematic errors are included).
The data are in
good agreement with the NLO TR-VFS QCD calculation using MRST99
(extended) PDFs (dashed line).
The solid line is a leading order CCFR QCD inspired fit
used for acceptance and radiative corrections.
A disagreement between the  CCFR data and CDHSW data
is observed in the slope of the $y$ distribution
at small $x$, and in the level of the cross sections at large $x$.}
\label{fig:diff1}
\end{figure}

The CCFR experiment  collected  data
using the Fermilab Tevatron Quad-Triplet wide-band  $\nu_\mu$ and $\nub_\mu$
beam. The CCFR detector~\cite{CALIB} consists of a
     steel-scintillator target calorimeter
instrumented with drift chambers, followed by a toroidally
magnetized muon spectrometer.
The hadron energy resolution is
$\Delta E_h/E_h = 0.85/\sqrt{E_h}$(GeV), and the muon momentum resolution is
$\Delta p_\mu/p_\mu = 0.11$. By measuring the hadronic energy ($E_h$), muon
momentum ($p_\mu$), and muon angle ($\theta_\mu$), we construct
three independent kinematic variables $x$, $Q^2$, and $y$.
The relative flux at different energies, obtained from the events
with low hadron energy ($E_h < 20$ GeV), is normalized so that
the neutrino total cross section equals the world average $\sigma^{\nu N}/E=
(0.677\pm0.014)\times10^{-38}$ cm$^2$/GeV and $\sigma^{\overline{\nu} N}
/\sigma^{\nu N}=0.499\pm0.005$~\cite{SEL}.
     After fiducial and kinematic cuts ($p_{\mu}>15$ GeV,
$\theta_{\mu} <0.150$, $E_h > 10$ GeV, and 30 GeV $<E_{\nu}<$ 360 GeV),
the data sample used for the extraction of structure functions consists of
1,030,000 $\nu_{\mu}$ and 179,000 $\nub_{\mu}$
events. Dimuon events are removed because of the ambiguous identification
of the leading muon for high-$y$ events.

The raw differential cross sections per nucleon on iron
     are determined in bins of $x$, $y$,
and $E_{\nu}$ ($0.01 < x < 0.65$, $0.05<y<0.95$, and $30< E_\nu <360$ GeV).
Over the entire $x$ region, differential cross sections are in
good agreement with NLO QCD calculation
using the Thorne and Roberts Variable Flavor Scheme
(TR-VFS)~\cite{MFS} with MRST99~\cite{MRST} extended~\cite{ext} PDFs
(with $R=R_{eff}^{\nu}$).
     This calculation includes an improved treatment of massive charm
production. The QCD predictions, which are on free neutrons and
protons, are corrected for nuclear~\cite{NUC},
higher twist~\cite{dupaper,YANGR}
and radiative effects~\cite{BARDIN}.

Figure \ref{fig:diff1}
shows some bins of the  differential cross sections
extracted at $E_\nu=85$ GeV
(complete tables of the differential cross sections
at all other energy bins are available~\cite{WWW}).
Also shown are the prediction of the NLO QCD TR-VFS calculation
using extended MRST99 PDFs, and
the prediction from a CCFR
     leading order Buras-Gaemers (LO-BG) QCD inspired fit~\cite{WWW}
used for calculation of acceptance and resolution smearing corrections
(uncertainties in these corrections are included in the sytematic
errors).
As expected from the quark parton model and QCD,
the CCFR data exhibit a quadratic $y$ dependence at small $x$
for $\nu_\mu$ and $\nub_\mu$, and a flat $y$ distribution at high $x$
for the $\nu_\mu$ cross sections.
Also shown are differential cross sections reported by the
CDHSW~\cite{CDHS}  collaboration.
A disagreement between the CCFR data and CDHSW data
is observed in the slope of the $y$ distribution
at small $x$, and in the level of the cross sections at large $x$.
This difference
is crucial in any QCD analysis which uses the CDHSW data.
For example, at the lowest $x$ bin the CDHSW $\nu_\mu$-Fe data continues
to increase with $y$,
in contrast to the small
decrease at large $y$ which is expected from the antiquark
component in the nucleon. In addition,
at the highest value of $x$ ( $x=0.65$), the level
of CDHSW $\nu_\mu$-Fe data does not agree
with CCFR or with the QCD predictions.
A recent QCD analysis~\cite{SBAR}  which includes these CDHSW data,
extracts an anomalously large asymmetry
between the $s$ and $\sbar$ quark distribution at high $x$ from
the CDHSW data. Since the $u$ and $d$ quark distributions are
very well constrained at this value of $x$ (from muon data on hydrogen
and deuterium), the only way to accommodate the high $x$ CDHSW data is
by the introduction of an asymmetric strange sea at high $x$.
The CCFR
data do not show this anomaly.

%

\begin{figure}[t]
\centerline{\psfig{figure=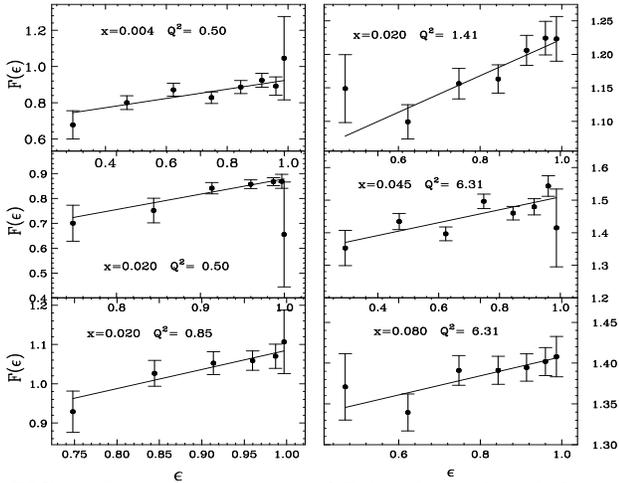,width=3.2in,height=2.5in}}
\caption{Typical extractions of $R$ (or $F_L$)
and $2xF_1$ for representative values of  $x$ and $Q^2$.}
\label{fig:linear}
\end{figure}

The raw differential cross sections are corrected for electroweak radiative
effects~\cite{BARDIN}, the $W$ boson propagator, and for the
     5.67\% non-isoscalar excess
of neutrons over protons in iron (only important at high $x$).
Values of $R$ (or equivalently $F_L$)
and $2xF_1$ are extracted from the sums of
the corrected $\nu_\mu$-Fe and $\nub_\mu$-Fe
differential cross sections at different
energy bins according to Eq. (1).
An extraction of $R$ using Eq. (1) requires a knowledge of
$\Delta xF_3$ term.
We obtain
$\Delta xF_3$ from  theoretical predictions for massive
charm production using the TR-VFS NLO calculation
with the extended MRST99 and the suggested scale $\mu=Q$.
This prediction is used as input to Eq. (1) in the extraction of $R^\nu$.
This model yields $\Delta xF_3$ values similar
to the NLO ACOT Variable Flavor Scheme\cite{ACOT},
(implemented with CTEQ4HQ~\cite{cteq4hq}
and the recent ACOT~\cite{vfs} suggested scale $\mu = m_c$ for $Q<m_c$, and
$\mu^2=m_c{^2}+0.5Q^2(1-m_c{^2}/Q^2)^n$ for $Q>m_c$ with $n=2$).
A discussion of the various theoretical calculations for $\Delta xF_3$
can be found in references~\cite{DXF3,dxf3_pred}.
Because of the positive correlation between $R$ and  $\Delta xF_3$,
the uncertainty in $\Delta xF_3$ introduces a model systematic error
at low $x$. However,  for $x>0.1$,  the $\Delta xF_3$ term
is small,
and the extracted values of $R^\nu$ are not sensitive to $\Delta xF_3$.
For the systematic error on the assumed level of
$\Delta xF_3$, we vary the strange sea and charm sea simultaneously by
$\pm 50$ \% ($\Delta xF_3$ is directly sensitive to the strange sea minus
charm sea). Note that the extracted value of $R$ is larger for a larger
input $\Delta xF_3$ (i.e. a larger strange sea).

Figure \ref{fig:linear} shows typical extractions of $R$ (or $F_L$)
and $2xF_1$ for a few
values of $x$ and $Q^2$. The extracted
values of $R^\nu$ are sensitive to the energy
dependence of the neutrino flux ($\sim$ $y$ dependence),
but are insensitive to the absolute normalization.
The uncertainty on the flux shape is estimated by
constraining
$F_2$ and $xF_3$ to be flat over $y$ (or $E_{\nu}$) for each $x$ and $Q^2$ bin.

\begin{figure}[t]
\centerline{\psfig{figure=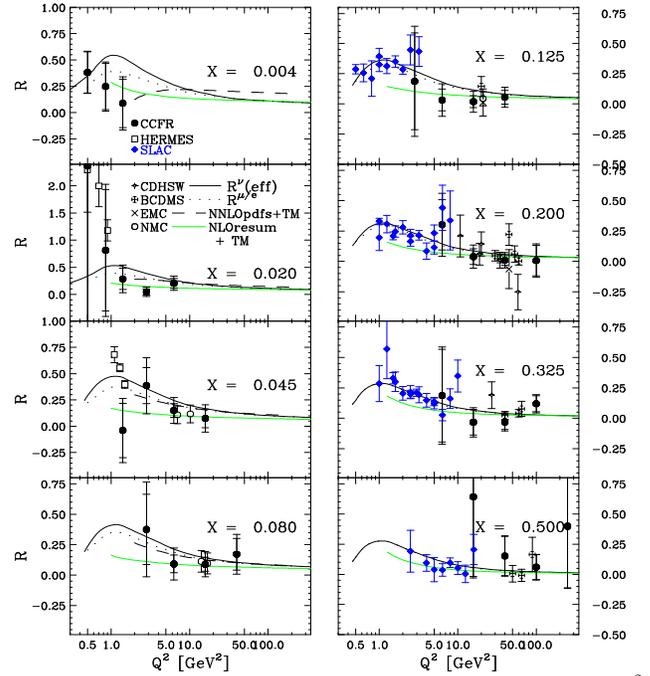,width=3.2in,height=3.5in}}
\caption{CCFR measurements of $R^\nu$ as a function of $Q^2$ for fixed $x$,
compared with electron and muon data, with the $R_{world}^{\mu/e}$ and
$R_{eff}^{\nu}$ ($m_c=1.3$) fits, with $R_{NNLOpdfs+TM}^{\mu/e}$
QCD calculation including NNLO PDFs (dashed),
and with $R_{NLOresum+TM}^{\mu/e}$ (dotted).
The inner errors include both statistical and experimental
systematic errors  added in quadrature. The outer errors include
the additional $\Delta xF_3$ model errors (added linearly).
Also shown are the HERMES results for $R_{N14}^{e}$
at small $x$ and $Q^2$.}
\label{fig:R}
\end{figure}

The extracted values of $R^\nu$ are
shown in Fig.~\ref{fig:R} for fixed $x$ versus $Q^2$.
The inner errors include both statistical and experimental
systematic errors (of similar magnitude on average~\cite{WWW})
added in quadrature. The the outer errors include
the additional $\Delta xF_3$ model errors (added linearly).
At the very lowest $Q^2$ values, the model
error is reduced because
all models for $\Delta xF_3$ approach zero around $Q^2 =0.4$ GeV$^2$.
This is because the strange quark distribution is expected to approach zero
for $Q$ values close to twice the mass of the strange quark. In addition,
the very low $Q^2$ region is below charm
production threshold. Note that the very low $Q^2$ and
low $x$ region is
of interest because it is where HERMES reports~\cite{HERMES} an
anomalous increase in $R^e$ for nuclear targets.

\begin{figure}[t]
\centerline{\psfig{figure=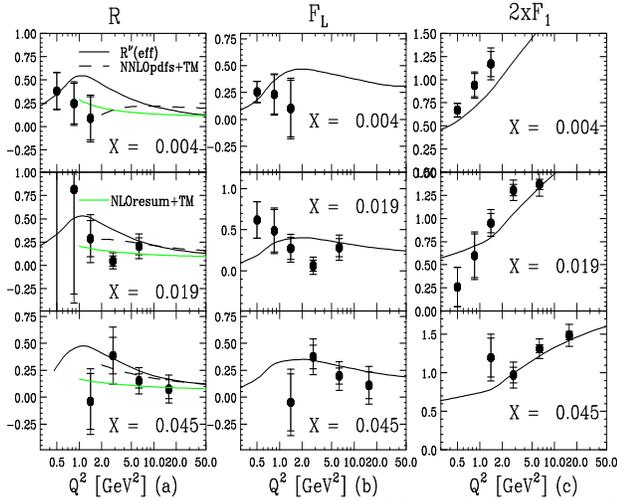,width=3.2in,height=2.6in}}
\caption{CCFR measurements of $R$~(a), $F_L$~(b) and $2xF_1$~(c)
data as a function of $Q^2$ for $x<0.05$.
The curves are the predictions from a QCD inspired leading
order fit to the CCFR differential cross section
data $R$= $R_{eff}^{\nu}$. Also shown is the $R_{NNLOpdfs+TM}^{\mu/e}$
QCD calculation including NNLO PDFs (dashed) and
$R_{NLOresum+TM}^{\mu/e}$ (dotted).}
\label{fig:FL_2xF1}
\end{figure}

The CCFR  $R^\nu$ values
are in agreement with measurements
of $R^{\mu/e}$~\cite{RWORLD,NMCR,R1998,slacr,otherr}, and
     also in agreement with both
the  $R_{world}^{\mu/e}$ and $R_{eff}^{\nu}$ fits.
At low $x$,  the data are lower than
the extrapolated values from these two fits.
Also shown in Fig.~\ref{fig:R}
and Fig.~\ref{fig:FL_2xF1} are the two most recent
calculations $R_{NNLOpdfs+TM}^{\mu/e}$ (dashed line)
and $R_{NLOresum+TM}^{\mu/e}$ (dotted line).
Note that a complete calculation
should include both the NNLO and the  $ln(1/x)$ resummation terms.
The calculations including either of these higher
order terms yield values of $R$  at small $x$ and
low $Q^2$ which are lower than $R_{world}^{\mu/e}$, and
are in better agreement with the data.
However, at low $x$ for $Q^2<5$ GeV$^2$ there are
large uncertainties in $R_{NNLOpdfs+TM}^{\mu/e}$
(mostly from the NNLO gluon distribution).

Also shown are the HERMES electron scattering results in nitrogen
at low values $x$.
The HERMES data~\cite{HERMES} for $R$ are extracted
from their ratios for $R_{N14}/R_{1998}$
by multiplying by the values from the $R_{1998}$ fit~\cite{R1998}.
The CCFR data do not clearly show a large anomalous increase at very
low $Q^2$ and low $x$.
It is expected that any nuclear effect
in $R$ would be enhanced in the CCFR iron target
with respect to the nitrogen target in HERMES.  However,
depending on the origin, the effects in electron versus $\nu_\mu$ charged
current scattering could be different.

The CCFR measurements of $F_L$ and $2xF_1$
     as a function of $Q^2$ for $x<0.05$ are shown in
Fig.~\ref{fig:FL_2xF1}.
The curves are the predictions from a QCD inspired leading
order fit to the CCFR differential cross
section data with $R$= $R_{eff}^{\nu}$. The extracted values at
the very lowest $x$ and $Q^2$ do not show any anomalous increase in
$R$ in our iron target. At the lowest values of  $x$, the disagreement
between the QCD inspired fit and the data is because
$R$= $R_{eff}^{\nu}$ was assumed (but our data and the most
recent theoretical calculations favor smaller values
of $R$ in this region).

In conclusion,
over the $x$ and $Q^2$ range where perturbative QCD is
expected to valid, $R^\nu$
is in good agreement with $R^{\mu/e}$
     data, and with the NNLO QCD calculation including NNLO PDFs and target
mass effects.
A very large nuclear enhancement in $R$ (as
reported by the HERMES experiment for electron scattering
on nitrogen) is not clearly observed in $\nu_\mu$-Fe scattering.
A comparison between CCFR and CDHSW differential cross section
indicates that although the cross sections agree over most
of the kinematic range, the CCFR data do not show the deviations
from the QCD expectations that are seen in the CDHSW data at
very low and very high $x$.


\begin{references}
\bibitem{NUC}
We use a fit to all $\mu$ and $e$ $F_2$ data on iron and deuterium.
\end{references}
\end{document}